\documentclass[aip,notitlepage, nofootinbib, floatfix, pre, reprint]{revtex4-1}%
\usepackage{amssymb}
\usepackage{amsfonts}
\usepackage{amsmath}
\usepackage{graphicx}
\usepackage{enumerate}

\usepackage[usenames,dvipsnames]{color}
\usepackage{ulem}%
\providecommand{\U}[1]{\protect\rule{.1in}{.1in}}

\begin{document}

\title{Unification of classical nucleation theories via unified It\^{o}-Stratonovich stochastic equation}

\author{Miguel A. Dur\'{a}n-Olivencia}
\email{m.duran-olivencia@imperial.ac.uk}
\altaffiliation[Present address: ]{Department of Chemical Engineering, Imperial College London, South Kensington Campus, London SW7 2AZ, UK}

\author{James F. Lutsko}
\email{jlutsko@ulb.ac.be}
\homepage{http://www.lutsko.com}

\affiliation{\label{CENOLI}Center for Nonlinear Phenomena and Complex Systems, Code Postal 231,
Universit\'{e} Libre de Bruxelles, Blvd. du Triomphe, 1050 Brussels, Belgium}

\begin{abstract}
Classical nucleation theory (CNT) is the most widely used framework to describe the early stage of first-order phase transitions. Unfortunately the different points of view adopted to derive it yield different kinetic equations for the probability density function, e.g. Zeldovich-Frenkel or Becker-D\"{o}ring-Tunitskii equations. Starting from a phenomenological stochastic differential equation a unified equation is obtained in this work. In other words, CNT expressions are recovered by selecting one or another stochastic
calculus. Moreover, it is shown that the unified CNT thus obtained produces the same Fokker-Planck equation as that from a recent update of CNT [J.F. Lutsko and M.A. Dur\'{a}n-Olivencia, {J. Chem. Phys.}, 2013, \textbf{138}, 244908] when mass transport is governed by diffusion. Finally, we derive a general induction-time expression along with specific approximations of it to be used under different scenarios. In particular, when the mass-transport mechanism is governed by direct impingement, volume diffusion, surface diffusion or interface transfer.
\end{abstract}
\date{\today }
\maketitle

\section{Introduction\label{introduction}}
Over the last century the classical approach to nucleation has reported an excellent ability to catch the essential rules underlying noise-induced phase transitions. The set of ideas which constitutes this framework were developed for over more than half a century. This development started with Gibbs' work \cite{article:gibbs-seminal-1878-a,article:gibbs-seminal-1878-b,book:gibbs-1931} on near-equilibrium phase transitions. Several years later, \citeauthor{article:volmer-weber-1926}\cite{article:volmer-weber-1926,book:volmer-1939} introduced kinetic aspects improving the purely thermodynamical picture given by Gibbs. This was further developed from a more atomistic point of view by \citet{article:farkas-1927} who developed Szilard's ideas and by \citet{article:becker-doring-1935}, \citet{article:tunitskii-1941},  \citeauthor{book:frenkel-1946}\cite{article:frenkel-1939,book:frenkel-1946} and
\citet{article:zeldovich-1943} within the context of liquid-vapor transitions. Finally, \citet{article:turnbull-fisher-1949} extended 
such a formalism with the aim of describing solid nucleation firstly from liquid and then from solid phases. Besides being intuitively appealing, classical nucleation theory (CNT) has shown an overwhelming robustness. Not only is it able to generate a satisfactory estimation for the nucleation rate equation which is good for practical purposes\cite{book:kashchiev-2000,book:kelton-2010}, but also it provides a natural mechanism for cluster formation which turns out being more real than initially expected\cite{article:sleutel-2014}. In more recent years, a massive amount of studies have renovated, if not improved, CNT based on either phenomenological or fundamental grounds\cite{article:shizgal-barret-1989,article:kashchiev-1969b,article:becker-doring-1935,
article:kaichew-stranski-1934,article:zeldovich-1943,
article:frenkel-1939,book:frenkel-1946,
article:tunitskii-1941,inbook:kelton-1991,article:lutsko-2011-b,article:prestipino-2012,article:lechner-2011,article:langer-1967,article:langer-1969,
article:lothe-1966,article:reguera-1998, article:mcgraw-2001,article:lutsko-2011-c,article:lutsko-2012-dtn,
article:lutsko-2012-a1,article:duran-otalora-2013,article:lutsko-duran-2013,article:duran-lutsko-2015}. 
The primary goal of this paper is to describe a systematic method to derive the Fokker-Planck equation for the probability to find clusters of a given size within the general setting of CNT. This results into a surprising generalization of the usual form of the stationary distribution compared to what is usually assumed in the classical theory: rather than a simple Boltzmann distribution depending on the work of formation of a cluster, there is a state-dependent prefactor that depends on the chosen stochastic calculus. For the particular choice of the Stratonovich calculus, the resulting expression reproduces the one recently derived from a more elaborate formalism based on fluctuating hydrodynamics\cite{article:lutsko-2012-dtn,article:lutsko-duran-2013}. Thus, one of the main contributions of this work is to provide a simpler route to that result based solely on CNT. From that, another important contribution comes up with the derivation of a general induction-time expression accompanied with specific approximations for it to be used in a variety of scenarios. Special attention is paid to those where the mass-transport mechanism is governed by direct impingement, volume diffusion, surface diffusion or interface transfer. To do this we followed the scheme outlined below.

In section \ref{theory} a stochastic differential equation (SDE) is proposed to model the time-evolution equation for the cluster size under a general stochastic calculus which is parametrized by $\alpha$ so that $\alpha=0,\frac{1}{2}$ and $1$ corresponds to  It\^{o}, Stratonovich and backward-It\^{o} conventions\cite{book:risken-1996a,book:oksendal-2003}, respectively. The state-dependent components of the postulated SDE are derived from a phenomenological point of view, following the CNT reasoning. The Fokker-Planck equation connected to such a SDE is then introduced in section \ref{fokker-planck}, being now apparent the similarities with the Zeldovich-Frenkel equation, except for presenting an additional term which affects the effective energy barrier. The effects of such a term are qualitatively explained in the same section. While the Zeldovich-Frenkel equation is recovered under backward-It\^{o} convention, the one derived by \citeauthor{article:lutsko-2012-dtn}\cite{article:lutsko-2012-dtn,article:lutsko-duran-2013} is 
 obtained under Stratonovich calculus. Lastly, the It\^{o} convention is studied which unveils the necessary condition for the different Fokker-Planck equations (FPEs) to converge. Section \ref{induction-time} is devoted to the derivation of a generalized expression for the mean first-passage time as well as different approximations for the most common experimental conditions. Finally our results are summarized in section \ref{conclusions}.

\section{Theory\label{theory}}

Nucleation is a quite complex  process which certainly requires a huge amount of relevant variables (order parameters) to be fully described. However, as we commented in section \ref{introduction}, CNT is based on a manageable description of the process which considers a single order parameter, namely the size of the emerging embryo (cluster) of the new phase within the old one. This is commonly assumed to be spherical and to have the same physical properties as those of the final stable state\cite{book:kashchiev-2000,book:kelton-2010}. This apparently crude simplification is nevertheless more than efficient when it comes to predicting the functional dependence of the nucleation rate on the thermodynamic quantities involved\cite{book:kashchiev-2000,book:kelton-2010}. That is why our study starts with a proposed SDE for the time evolution of the cluster size. The reasons why we are motivated to propose such an equation are mainly two. On the one hand, it is known that nucleation is a thermally activated process involving the escape from a metastable state via overcoming an energy barrier\cite{book:kashchiev-2000,book:kelton-2010,book:barrat-hansen-2003}. Thus, if the nucleation process is governed by a single order parameter it should undergo, at least in good approximation, a Langevin equation as proposed by Kramers\cite{article:hanggi-1990}.
On the other hand, recent studies\cite{article:lutsko-2012-dtn,article:lutsko-duran-2013,article:duran-lutsko-2015} have shown a formal derivation of such kind of SDEs when the mass transport is governed by diffusion. That said, a general Langevin equation is assumed and their drift and diffusion terms will be deduced from phenomenological arguments, and inspired by the ideas underlying CNT. Without further ado, we will proceed writing this up.

\subsection{Stochastic dynamics}

Let us consider a system which is set in a metastable state and direct our attention to an arbitrary spherical cluster of those  randomly growing and shrinking. Let $X$ be the size of such a cluster, accounting for the number of molecules inside it. Following Kramers' reasoning for thermally-activated escape processes\cite{article:hanggi-1990} we will propose the model equation,
\begin{equation}
 dX(t)=\eta^{-1}(X,t)F(X,t)dt+\sqrt{2k_BT\eta^{-1}(X,t)}\star dB(t)\label{eq:1}
\end{equation}
with $\eta$ being an effective viscosity, $F$ being the effective force acting on $X$, $T$ is the temperature, $k_B$ the Boltzmann constant and $dB(t)$ being a Wiener process. Intrinsic to this equation is the consideration of $X$ as a continuous variable, approach first introduced by \citeauthor{article:zeldovich-1943}\cite{article:zeldovich-1943,book:kashchiev-2000,book:zettlemoyer, book:dunning-zettlemoyer-1969, book:walton-zettlemoyer-1969,book:abraham-1974}. Based on the work of \citet{chapter:ree-ree-ree-eyring-1962} it was shown (e.g. Sec. 9.1 of Ref. \onlinecite{book:kashchiev-2000}) that clusters of more than a few molecules can be fairly well described within the framework of the continuous approach, and so will we consider it as a good enough paradigm. The star product, $\star$, was introduced to remark that we are using a general stochastic calculus (hereafter called $\alpha$-calculus) defined by means of the definition of the stochastic integral,\cite{article:coutinho-2010,article:brettschneider-volpe,article:lau-2007,article:lancon-2002}
\begin{align}
I_\alpha[R(x,t)]=&\int_{t_0}^{t}dB(t') R(x(t'),t')\nonumber\\
:=&\text{ms}\,\,\text{-}\lim_{n\rightarrow\infty}\sum_{i=0}^{n-1}R(x(t_i^*),t_i^*)\Delta B_i\label{eq:2}
\end{align}
where $R(x,t)$ is a left-continuous function, i.e. a function
which is continuous from the left at all the points where it is
defined, the symbol ``$:=$'' means definition, (ms-lim) represents \emph{mean squared} limit, i.e. a second moment convergence, $t_i^*=(1-\alpha)t_i+\alpha t_{i+1}$ and $\Delta B_i=B(t_{i+1})-B(t_i)$. It can be shown\cite{article:coutinho-2010,article:brettschneider-volpe,article:lau-2007,article:lancon-2002} that choosing $\alpha=0,\frac{1}{2}$ or $1$, one recovers It\^{o}, Stratonovich or backward-It\^{o}'s definition. At the moment we will focus our efforts on obtaining a good estimation of the drift term and later on we will discuss the consequences of selecting one or another value of $\alpha$. 

As previously mentioned, our aim is providing an alternative route of derivation of CNT starting from the cluster-growth law. Based on CNT, we know that the effective force, $F$, is related to the \emph{work of cluster formation}, $W$, through its derivative,
\begin{equation}
 F(X,t)=-\frac{\partial W(X,t)}{\partial X}\equiv - W'(X,t).
 \label{eq:3}
\end{equation}
The work of cluster formation is usually expressed as the increment of free energy experienced by the system due to the emergence of a cluster of size $X$. Depending on the system under consideration that work is specified in terms of either the Gibbs ($\Delta\mathcal{G}$) or Helmholtz ($\Delta\mathcal{F}$) free energy, or the Grand (Landau) potential ($\Delta\Omega$). In spite of making the derivation as general as possible, we do not specify a given thermodynamic potential since knowing the functional dependence of $F$ on $W$ is enough by far.

The derivation of an expression for $\eta$ requires nonetheless a slightly longer discussion. According to CNT, the effective time a cluster will spend to lose a molecule is given by the inverse of the difference between monomer attachment, $f$, and detachment, $g$, frequencies,
\begin{equation}
 \tau_\leftarrow  = \frac{1}{g(X,t)-f(X,t)}.
\label{eq:4}
\end{equation}
Within nucleation regime, clusters experience a stronger force to shrink than to grow due to their metastable nature. This results in a higher detachment rate than the corresponding attachment frequency (Chap. 10 or Ref. \onlinecite{book:kashchiev-2000}) with $\tau_\leftarrow\ll 1\,\text{s}$, typically of order  $10^{-7}-10^{-12}\,\text{s}$ as discussed by \citeauthor{book:kashchiev-2000} (Figs. 10.2, 10.4 and 10.6 of Ref. \onlinecite{book:kashchiev-2000}).

On the other hand, from the definition of the Kramers-Moyal coefficients\cite{book:risken-1996a} 
\begin{equation}
  D^{(n)}(x,t)=\frac{1}{n!}\lim_{\tau\rightarrow0}\left.\frac{1}{\tau}\langle[X(t+\tau)-x]^n\rangle\right|_{X(t)=x},
\label{eq:5}
\end{equation}
it was argued (Refs. \onlinecite{article:brettschneider-volpe} and \onlinecite{article:lancon-2002}) that the first Kramers-Moyal coefficient related to equation (\ref{eq:1}) is given by,
\begin{align}
D^{(1)}(X,t)=&\, a(X, t) + \alpha\,\frac{\partial b(X,t)}{\partial X}b(X,t)\label{eq:6}\\
=&-\frac{1}{\beta\eta(X,t)}\left(\beta W'(X,t)+\alpha\frac{\partial}{\partial X}\ln \eta(X,t)\right),\notag
\end{align}
with  $a(X,t) = \eta^{-1}(X,t)F(X,t)$ and $b(X,t) = \sqrt{2k_BT\eta^{-1}(X,t)}$ being the drift and diffusion forces, and $\beta=1/k_BT$. Now, we can use the fact that the time $\tau_\leftarrow$ is expected to be small compared to the typical time scale associated with a significant change of $\langle X(t) \rangle$ so that the limit in Eq.(\ref{eq:5}) can be approximated by evaluation at $\tau = \tau_{\leftarrow}$ for which $\langle X(t+\tau_{\leftarrow})-X(t)\rangle=-1$, giving
\begin{equation}
 D^{(1)}(X,t)=\left\langle\frac{dX}{dt}\right\rangle\sim f(X,t)-g(X,t)
 \label{eq:7}
\end{equation}
which agrees with the cluster-growth law derived in CNT,\cite{book:kelton-2010}
\begin{equation}
\left\langle\frac{dX}{dt}\right\rangle_{\text{CNT}}=f(X,t)-g(X,t)
 \label{eq:8}
\end{equation}
As we know from CNT, whereas the analytical expression of $f(X,t)$ can be derived from collision theory\cite{book:kashchiev-2000,book:kelton-2010}, finding the frequency $g(X,t)$ is not a trivial task. This is so because monomer detachment depends on parameters characterizing the cluster which may differ appreciably from those of the bulk new phase. To get rid off this problematic quantity we will follow the same reasoning as in CNT.

Let us assume that $f(X,t)$ is a well determined quantity. We can evaluate the difference $f(X,t)-g(X,t)$ in terms of $f(X,t)$ making use of the detailed balance equation,
\begin{equation}
 f(X-1,t)\,\widetilde{P}(X-1,t)=g(X,t)\,\widetilde{P}(X,t)
\label{eq:9}
\end{equation}
with $\widetilde{P}(X,t)$ being the {quasi-equilibrium probability density function}\cite{article:kashchiev-1},
\begin{equation}
 \widetilde{P}(X,t)=P_0(t)\sigma(X,\alpha;t)\,e^{-\beta W(X,t)}
\label{eq:10}
\end{equation}
where $P_0(t)$ is an instantaneous normalization constant and $\sigma$ is some function of $X,\,\alpha$ and $t$. Looking to accomplish our goal, we will follow Kashchiev\cite{article:kashchiev-1} and approximate $f(X-1,t)\widetilde{P}({X-1},t)$ by the truncated Taylor expansion about point $X$,
\begin{align}
 f(X-1,t)\widetilde{P}&(X-1,t)\notag\\
 \sim&f(X,t)\widetilde{P}(X,t)-\frac{\partial}{\partial X}f(X,t)\widetilde{P}(X,t),
\label{eq:11}
\end{align}
and, hence,
\begin{align}
 f(X,t)-g(X,t)\sim& f(X,t)\frac{\partial}{\partial X}\ln f(X,t)\widetilde{P}(X,t).
 \label{eq:12}
\end{align}
Using equation (\ref{eq:10}) into (\ref{eq:12}) we get,
\begin{align}
 f(X,t)-g(X,t) =& -f(X,t)\frac{W'(X,t)}{k_BT} \label{eq:13}\\
 &+f(X,t)\frac{\partial}{\partial X}\ln \left[\sigma(X,\alpha;t)f(X,t)\right], \notag
\end{align}
where we replaced ``$\sim$'' by ``$=$'' assuming that these approximations are accurate enough for all practical purposes.
Finally, by substituting equation (\ref{eq:13}) into (\ref{eq:7}) and equating to equation (\ref{eq:6})
we eventually get,
\begin{align}
 \eta^{-1}(X,t)&= \beta {f(X,t)},\label{eq:14}\\
 \sigma(X,\alpha;t)&=f^{\alpha-1}(X,t)\notag.
\end{align}

Having equations (\ref{eq:3}) and (\ref{eq:14}), the Langevin equation proposed at the very beginning (Eq. \ref{eq:1}) can be finally rewritten,
\begin{equation}
  dX(t)=-f(X,t)\frac{\partial\beta W(X,t)}{\partial X}dt+\sqrt{2 f(X,t)}\star dB(t)\label{eq:15}.
\end{equation}
This stochastic equation is equivalent to that derived by \citeauthor{article:lutsko-2012-dtn}\cite{article:lutsko-2012-dtn,article:lutsko-duran-2013} from fluctuating hydrodynamics for a single-order parameter, when we set $\alpha=\frac{1}{2}$ (i.e. Stratonovich's calculus). Moreover, as it will be checked in section \ref{fokker-planck} this equation is statistically equivalent to the Zeldovich-Frenkel equation for $\alpha=1$ (i.e. backward-It\^{o}'s calculus), given that it produces the same time-evolution equation for the probability density function.  It is thus interesting to see how the hypotheses underlying Zeldovich's derivation are equivalent to choose a specific stochastic calculus. Besides, now the equilibrium with the thermal bath is always ensured regardless the value of $\alpha$, unlike the common belief that only the backward-It\^{o} convention is capable of guaranteeing equilibrium\cite{article:lau-2007}. Actually, such a belief is reached after considering the equilibrium regime as equivalent to a Boltzmann distribution law. Nevertheless from previous studies we know this is not true for nucleation, where the equilibrium distribution derived from a fluctuating-hydrodynamic framework shows a state-dependent prefactor.\cite{article:lutsko-2012-dtn,article:lutsko-duran-2013}
That is why we free the derivation herein presented of that restriction assuming instead a local-equilibrium law (Eq. \ref{eq:10}) with a general state-dependent prefactor. This results in a general equilibrium distribution function (using Eqs. (\ref{eq:10}) and (\ref{eq:14})) which yields a state-independent pre-exponential factor for $\alpha=1$, as expected. Noteworthy in such a case the resulting theory cannot be covariant as will be shown in section \ref{sec:eqAndStDists}.

Nevertheless, equation (\ref{eq:15}) differs from that produced by  \citeauthor{article:tunitskii-1941}'s equation\cite{article:tunitskii-1941,chapter:penrose-2008} 
\begin{equation}
 dX(t)=\left(f(X,t)-g(X,t)\right)dt+\sqrt{f(X,t)+g(X,t)}\,dB(t)\label{eq:16}
\end{equation}
which is interpreted under It\^{o}'s convention, i.e. $\alpha=0$. In order to know under which circumstances both equations are equivalents, we will follow the same reasoning as that led us to equation (\ref{eq:14}), and so we arrive at,
\begin{align}
 f(X,t)-g(X,t)=&\ -{W'(X,t)}\,\eta^{-1}(X,t),\label{eq:17}\\
 f(X,t)+g(X,t)=&\ {2k_BT} \eta^{-1}(X,t)\label{eq:18}.
\end{align}
Then,
\begin{align}
 \frac{g(X,t)}{f(X,t)}=&1+\frac{1}{1-\left(\frac{W'(X,t)}{2k_BT}\right)}\frac{W'(X,t)}{k_BT}
\label{eq:19}
\end{align}
To get the usual approximation of $g(X,t)$ given in CNT (e.g. Eq. (10.90) of Ref. \onlinecite{book:kashchiev-2000}), i.e.,
\begin{equation}
 g(X,t)=f(X,t)\exp\left(\frac{\partial \beta W(X,t)}{\partial X}\right)\label{eq:20}
\end{equation}
we need
\begin{equation}
 \frac{\partial\beta W(X,t)}{\partial X}\ll 1\label{eq:21}.
\end{equation}
given that in this limit one gets,
\begin{align}
 \frac{g(X,t)}{f(X,t)}\sim&1+\frac{\partial \beta W(X,t)}{X}+\frac{1}{2}\left(\frac{\partial\beta W(X,t)}{\partial X}\right)^2\nonumber\\
\sim&\exp\left(\frac{\partial \beta W(X,t)}{\partial X}\right)\label{eq:22}
\end{align}
That way, our proposed model will recover the Tunitskii equation under It\^{o}'s convention for slowly-varying energy barriers, which indeed agrees with the hypotheses underlying CNT.

Thus far, we have derived heuristically a model (Eq. \ref{eq:15}) which seems to be in accordance even with more rigorous and modern theories. In what follows, we will study the dynamics of the probability distribution function (PDF) associated with equation (\ref{eq:15}) in order to make contact with CNT. This will make possible to get some important quantities such as the stationary distribution function (for undersaturated systems) or the nucleation (or induction) time.

\section{Fokker-Planck equation\label{fokker-planck}}

The time-evolution equation of the PDF, $P(X,t)$, of the random variable $X$ will be given by the the following Fokker-Planck equation:\cite{article:coutinho-2010,article:brettschneider-volpe,article:lau-2007}
\begin{align}
  \frac{\partial P(X,t)}{\partial t}=-\frac{\partial \mathfrak{J}(X,t)}{\partial X}, \label{eq:23}
\end{align}
with 
\begin{align}
 \mathfrak{J}(X,t)=&-\left\{
  f(X,t)\frac{\partial }{\partial X}\left[
\beta W(X,t)+(1-\alpha)\ln f(X,t)\right] \right.\notag\\
  &\left.
  +\,f(X,t)\frac{\partial }{\partial X}
  \right\}
  P(X,t).
  \label{eq:24}\\
  =&-\left(
  f(X,t)\frac{\partial \beta\Phi(X,t)}{\partial X}+f(X,t)\frac{\partial }{\partial X}
  \right)
  P(X,t),\label{eq:25}
\end{align}
with, 
\begin{equation}
\beta\Phi(X,t)= \beta W(X,t)+(1-\alpha)\ln f(X,t) .
\label{eq:26}
\end{equation}
Now, the similarities between this FPE and that obtained in CNT are apparent. Yet more, the Zeldovich-Frenkel equation is recovered when the backward-It\^{o} convention is adopted. Surprisingly enough, this naive model also recovers the FPE given in more recent rederivations of CNT\cite{article:lutsko-2012-dtn,article:lutsko-duran-2013} when the Stratonovich calculus is considered.

\subsection{Short-time propagators: critical clusters with growing habits}

Now we are going to evaluate the impact of the extra logarithmic term. Since $0\leq \alpha\leq 1$, it is evident that it entails an increase in the energy barrier with respect to the Zeldovich-Frenkel equation. However, there is another interesting effect coming from this additional term which has to do with the probability of a critical cluster (defined by $\beta W'(X_*,t)=0$) to grow or shrink. While it is customary accepted that these probabilities must be the same, this is only true under It\^{o}'s convention, as we will show right away. To this end, we will make use of short-time propagators\cite{book:risken-1996a}.

It is known from collision theory that the analytical equation of $f(X,t)$ is size-dependent and so is $\sqrt{2 f(X,t)}$. Hence, their values will change from the initial to the final size during a unitary jump in the size axis, $X$. This change implies that the cluster feels different attachment rates in going from an initial size, $X_0$, to another, $X$. Here is where the choice of the stochastic calculus comes into play, since each of them corresponds to a different origin where evaluating the noise amplitude, which yields asymmetric probability distributions\cite{article:coutinho-2010,article:brettschneider-volpe} for any $\alpha\neq0$. Considering $\alpha\in[0,1]$ the short-time propagator\cite{book:risken-1996a} related to the FPE (\ref{eq:22}), and hence to the SDE (\ref{eq:15}), is given by  the following equation (\ref{appendix:a})
\begin{align}
 p_{\alpha}(X&,t+\tau|X_0,t)\notag\\
 =&\frac{
 \exp
\left(\begin{array}{l}
-\alpha\tau \frac{\partial D^{(1)}(\widetilde{X}_\alpha,t)}{\partial X}\\
\quad+\alpha^2\tau \frac{\partial^2 D^{(2)}(\widetilde{X}_\alpha,t)}{\partial X^2}\\
\qquad-\frac{\{ X-X_0-[D^{(1)}(\widetilde{X}_\alpha,t)-2\alpha D^{(2)'}(\widetilde{X}_\alpha,t)]\tau\}^2}{4\tau D^{(2)}(\widetilde{X}_\alpha,t)}
      \end{array}
\right)
 }{2\sqrt{\pi\tau D^{(2)}(\widetilde{X}_\alpha,t)}}
 ,\label{eq:27}
\end{align}
\begin{figure}[t]
\begin{center}
  \includegraphics{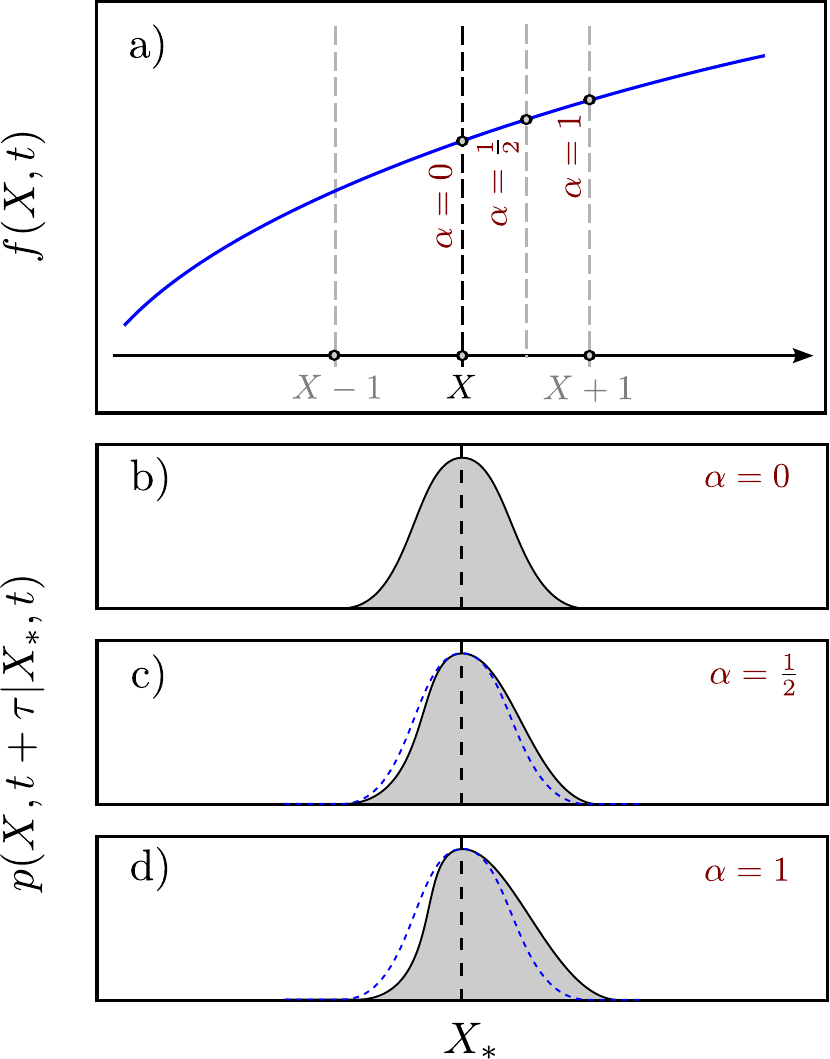}
\end{center}
\caption{
\label{fig:short-time-prop}
This figure is inspired by Fig. 4 of \citet{article:brettschneider-volpe} The top panel (a)
is a schematic representation of the dependence of the monomer attachment frequency on the cluster size. Below, panels (b), (c) and (d) show qualitatively the effect of the logarithmic term on the short-time propagators, in particular when the initial condition is the critical cluster. It is readily observed how the symmetry of growing or shrinking is broken for all $\alpha\neq0$.}
\end{figure}
\noindent{}with:
\begin{align}
D^{(1)}(\widetilde{X}_\alpha,t)=&\ -f(\widetilde{X}_\alpha,t)\beta W'(\widetilde{X}_\alpha,t)+\alpha f'(\widetilde{X}_\alpha,t)\notag,\\
D^{(2)}(\widetilde{X}_\alpha,t)=&\ f(\widetilde{X}_\alpha,t)\notag,\\
\widetilde{X}_\alpha=&\ \alpha X(t+\tau)+(1-\alpha)X(t)\notag\\
		    \equiv&\ \alpha X+(1-\alpha)X_0.\label{eq:28}
\end{align}
In the particular case of $X_0=X_*$ we know by definition that $\beta W'(X_*)=0$, i.e. the time evolution of the size for a critical cluster is purely stochastic (Eq. \ref{eq:15}). However, as we can observe in Fig. \ref{fig:short-time-prop}, the probability to grow only equals that to decrease for $\alpha=0$, given that
\begin{align}
 p_{\alpha=0}(X,t+\tau\,|&\,X_*,t)\notag\\
 =&\,\frac{1}{2\sqrt{\pi\,f(X_*,t)}}\exp\left(-\frac{(\Delta X_*)^2}{4f(X_*,t)\tau}\right)
 \label{eq:29},
\end{align}
where $\Delta X_*=X-X_*$. This is in line with the short-time propagator directly derived from Tunitskii's equation (Eq. \ref{eq:16}) when $f(X_*,t)=g(X_*,t)$, i.e. the probability to increase by one unit the cluster size must be equal to the probability to decrease by the same amount. 

Nevertheless, for all $\alpha\in(0,1]$ the short-time propagator gradually becomes an asymmetric distribution reaching the maximum deformation for\cite{article:brettschneider-volpe} $\alpha=1$,
\begin{align}
  p_{\alpha=1}&(X,t+\tau|X_*,t)\notag\\
  =&\,\frac{\exp\left(
  \begin{array}{l}
  \tau \left.\frac{\partial}{\partial x}\left(f(x,t)\frac{\partial\beta W'(x,t)}{\partial x}\right)\right|_{x=X}\\
  \quad-\frac{\{\Delta X_*+[f(X,t)\beta W'(X,t)+f'(X,t)]\tau\}^2}{4\tau f(X,t)}
  \end{array}\right)
  }{2\sqrt{\pi\,f(X,t)}}.
\label{eq:30}
\end{align}
Given that $f(X,t)$ is a monotonically increasing function of $X$ (e.g. Ch. 10 of Ref. \onlinecite{book:kashchiev-2000}), it can be shown that such an asymmetry favours the growth, instead of the shrinking. This fact has been already reported by \citet{article:brettschneider-volpe} in a different scenario but with similar conditions. Inspired by their discussion, this result has been schematically represented in Fig. \ref{fig:short-time-prop}. While this is a very interesting result which underlies the different derivations of CNT, it is true that for very large critical clusters, $X_*\gg1$, the general short-time propagator becomes,
\begin{align}
 p_{_{\text{CNT}}}(X,t+\tau|X_*,t)=&\lim_{X_*\gg1} p_\alpha(X,t+\tau\,|\,X_*,t)\notag\\
 \sim&\frac{\exp\left\{-\frac{(\Delta X_*)^2}{4f(X_*,t)\tau}\right\}}{2\sqrt{\pi\,f(X_*,t)}}\label{eq:31},
\end{align}
where we have considered the Szilard limit\cite{article:farkas-1927} in order to approximate $f(X,t)\sim f(X_*,t)$. Thus, we see that the different $\alpha$-calculi yield the same results only for very large critical clusters, which is equivalent to say for near-equilibrium systems. It is worth mentioning that a recent rederivation of CNT from fluctuating hydrodynamics\cite{article:lutsko-2012-dtn,article:lutsko-duran-2013,article:duran-lutsko-2015} (hereafter denoted dCNT) has formally shown that the right FPE is that given for $\alpha=\frac{1}{2}$, otherwise this will not be a covariant theory. Hence, from the above discussion, the critical cluster will experience a slightly higher tendency to grow, unlike what would be initially expected from CNT.

\subsection{Equilibrium and stationary distributions\label{sec:eqAndStDists}}

In this section we are going to explore the solutions of the  general FPE given by equation (\ref{eq:22}) and (\ref{eq:25}). It is widely known that finding the exact solution is a highly difficult problem, as well as potentially impossible. However, as discussed by \citet{article:hanggi-talkner-borkovec-1990}, 
an exact solution can be obtained assuming a stationary system with constant flux, $\mathfrak{J}_s$, which is ensured by removing clusters once they reach a given size, $X=X_{max}>X_*$ (see also Refs. \onlinecite{book:kashchiev-2000} and \onlinecite{book:lifshitz-theoretical-2010}). Then,  the steady-state distribution must satisfy $P_s(X_{max})=0$.
Let us consider $f(X,t)\equiv f(X)$ and $\Phi(X,t)\equiv \Phi(X)$. Coming back to equation (\ref{eq:25}),  
\begin{equation}
 \mathfrak{J}_s=-\left(
  f(X)\frac{\partial \beta\Phi(X)}{\partial X}+f(X)\frac{\partial }{\partial X}
  \right)
  P(X)
  \label{eq:32}
\end{equation}
which readily produces,\cite{book:risken-1996a}
\begin{align}
 P_s(X)=&A\,e^{-\beta\Phi(X)}-\mathfrak{J}_se^{-\beta\Phi(X)}\int^X\frac{e^{\beta\Phi(Y)}}{f(Y)}dY\label{eq:33}
\end{align}
with $A$ being a normalization constant. 
When the boundary condition on $P_s(X_{max})=0$ is imposed, Eq. (\ref{eq:33}) becomes, 
\begin{equation}
 P_s(X)=\mathfrak{J}_se^{-\beta\Phi(X)}\int_X^{X_{max}}\frac{e^{\beta\Phi(Y)}}{f(Y)}dY.
\end{equation}
Now, we can note that for some change of variable $X\rightarrow Y(X)$, the stationary probability should fulfill,
\begin{equation}
 \widetilde{P}_s(Y)=P_s(X)\left(\frac{dX}{dY}\right),
 \label{eq:34}
\end{equation}
which imposes the following condition on $f(X)$, 
\begin{equation}
 \widetilde{f}(Y)=f(X(Y))\,\left(\frac{dX}{dY}\right)^{1/(\alpha-1)}\label{eq:35}.
\end{equation}
Therefore, as we can check, the classical Zeldovich-Frenkel equation cannot be covariant\cite{article:lutsko-duran-2013} given the singularity occurring for $\alpha=1$. For all other values, whether the theory presents general covariance or not is conditioned by the definition of the attachment frequency. Thus far, the only derivation satisfying such a requirement has been dCNT within the context of diffusion-control mass transport\cite{article:lutsko-duran-2013}. As results from such a study, the effects related to the non-covariant character of CNT are subdominant.

If we consider an initial equilibrium (undersaturated) state, the stationary PDF becomes the equilibrium distribution by identifying it with a zero flux regime. Then, imposing $\mathfrak{J}_{s}=0$ into equation (\ref{eq:33}) we arrive at,
\begin{align}
 P_{eq}(X)=&\, A\,f^{(\alpha-1)}(X) e^{-\beta W(X)}
=\,A\,e^{-\beta\Phi(X)}
\label{eq:36}
\end{align}

The results introduced in this section will be remarkably important when it comes to the derivation of the induction times, characterized by the mean first-passage time (MFPT).

\subsection{Semiadiabatic limit}

Nevertheless, real experiments could involve time-dependent coefficients.  Following \citet{article:weidlich-haag-1980-a}, we will assume that such a dependence of both $f$ and $\Phi$ on time is controlled via a certain control function, $\kappa(t)$. Under this assumption we get,
\begin{align}
D^{(1)}(X,\kappa(t))=&-f(X,\kappa(t))\frac{\partial\beta W(X,\kappa(t))}{\partial X}\notag\\
&+\,\alpha \frac{\partial f(X,\kappa(t))}{\partial X},\notag\\
D^{(2)}(X,\kappa(t))=&\,f(X,\kappa(t)).\label{eq:37}
\end{align}
As evident, if $\kappa$ is time-independent we recover the previous results. In most cases, the control function will be either the average density of the metastable state, $\kappa(t)=\rho_{av}(t)$, or the temperature of the system, $\kappa(t)=T(t)$. In those cases in which $\kappa(t)$ is a slowly-varying function of time, with respect to a typical time scale, one expects that the stationary solution follows adiabatically the motion of $\kappa(t)$, i.e. the system reaches  a stationary state almost instantaneously. This hypothesis is also known as semiadiabatic limit\cite{article:weidlich-haag-1980-a}. The time scale that characterizes this limit can be interpreted as the relaxation time toward the initial metastable state, as pointed out by \citet{article:talkner-luczka}. Under these circumstances the zeroth order approximation for the quasi-stationary PDF is
\begin{align}
  P_{qs}(X;t)\sim&A(t)\,e^{-\beta\Phi(X,\kappa(t))}\notag\\
  &-\mathfrak{J}_se^{-\beta\Phi(X,\kappa(t))}\int^X\frac{e^{\beta\Phi(Y,\kappa(t))}}{f(Y,\kappa(t))}dY\label{eq:38},
\end{align}
and so, the quasi-equilibrium PDF for undersaturated conditions,  for which the flux necessarily vanishes ($\mathfrak{J}_s = 0$), is
\begin{align}
 P_{qe}(X;t)\sim&\,A(t)e^{-\beta\Phi(X,\kappa(t))}\label{eq:39}.
\end{align}
The semiadiabatic approach is quite useful in order to make a first approximation of the MFPT under non-stationary conditions. Nonetheless, for a more accurate approximation of induction times the path integral formalism developed by \citet{article:getfert-2010} should be considered.

\section{Estimation of induction times and nucleation rates\label{induction-time}}

In this section an approximation for the MFPT and so for the nucleation rate will be provided for most typical mechanisms governing mass transport, i.e. the attachment rate. To do that, we will firstly get an approximation of the MFPT and then we will particularize it by using the most used expressions for $f(X)$ in the literature\cite{book:kashchiev-2000}. The reason why we will focus on obtaining an approximation for the MFPT (hereafter denoted by $\tau$) is that this quantity is closely linked to the nucleation rate, $J$. Indeed, within the range of applicability of CNT one expects (e.g., Sec. 10.6 and Eq. 72 of Ref. \onlinecite{book:barrat-hansen-2003-a}),
\begin{equation}
 J_\alpha \sim \rho_{av}\tau^{-1}_{\alpha}.
 \label{eq:40}
\end{equation}
where the subscript $\alpha$ has been introduced to highlight the fact that it will depend on the $\alpha$-calculus selected.
That way, $\tau$ can be understood as the mean time required for nucleation to occur. For this purpose we will follow \citet{book:risken-1996a} to get an approximation of $\tau$. To begin with, we will consider stationary conditions, i.e. $f$ and $W$ time-independent. From the results thus obtained, a naive approximation for time-dependent conditions will be readily produced by considering the semiadiabatic limit. 
It is easy to show that equation (\ref{eq:32}) can be rewritten as,
\begin{align}
\mathfrak{J}_s=&-f(X)e^{-\beta\Phi(X)}\frac{\partial}{\partial X}\left(e^{\beta\Phi(X)}P(X)\right)\nonumber\\
 =&-e^{-\beta\varphi(X)}\frac{\partial}{\partial X}\left(e^{\beta\Phi(X)}P(X)\right)\label{eq:41},
\end{align}
with
\begin{equation}
 \varphi(X)=W(X)-\alpha\,k_BT\ln (f(X))\label{eq:42}.
\end{equation}
If the barrier is relatively  high, $\mathfrak{J}_s$ is expected to be very small. Hence, we can integrate equation  (\ref{eq:41}) from the minimum, $X=X_{min}$, to the maximum, $X=X_{max}$, size allowed for a cluster,
\begin{align}
 \mathfrak{J}_s\int_{X_{min}}^{X_{max}}&e^{\beta\varphi(s)}ds&\notag\\
 =&\ P(X_{min},t)e^{\beta\Phi(X_{min})}\left(1-\frac{P(X_{max})}{P(X_{min})}e^{\beta\Delta\Phi}\right)\nonumber\\
\sim&\ P(X_{min})e^{\beta\Phi(X_{min})}\label{eq:43}.
\end{align}
Under these conditions, as $\mathfrak{J}_s$ is assumed to be very small we can use Eq. (\ref{eq:33}) to approximate $P(X_{min})\sim A\,e^{-\beta\Phi(X_{min})}$ and $P(X) \sim A\,e^{-\beta\Phi(X)}$. That way, the distribution function near $X_{min}$ will be approximately given by (see Ref. \onlinecite{book:risken-1996a})
\begin{equation}
 P(X)\,\sim\, P(X_{min})e^{-\beta(\Phi(X)-\Phi(X_{min}))}\label{eq:44}.
\end{equation}
Then, we can get the following expression for the MFTP\cite{book:risken-1996a}
\begin{align}
 \tau(\alpha)\sim&\ \frac{P(X_{min},t)e^{\beta\Phi(X_{min})}\int_{X_{min}}^{X_{max}}e^{-\beta\Phi(s)}ds}{P(X_{min},t)e^{\beta\Phi(X_{min})} / \int_{X_{min}}^{X_{max}}e^{\beta\varphi(s)}ds}\nonumber\\
  \sim&\ \int_{X_{min}}^{X_{max}}e^{-\beta\Phi(s)}ds \int_{X_{min}}^{X_{max}}e^{\beta\varphi(s)}ds\label{eq:45}.
\end{align}
The usual procedure\cite{book:risken-1996a} is to expand both exponents around $X_{min}=0$ and $X_*$ respectively, but in this problem such a  method cannot be used, since $f(X)$ goes to zero as $X\rightarrow0$\cite{book:kashchiev-2000,book:kelton-2010}. Nonetheless, the main contribution to the first integral stems from the region around $X_{min}$. 
Thus, in the case of 3D nucleation the free energy term will be governed by the surface term\cite{book:kashchiev-2000} and hence, the exponent involving $\beta\Phi$ can be truncated in good approximation as below
\begin{equation}
\beta\widetilde{\Phi}_{3D}(X)= \beta\left(\Theta X^{2/3}+(1-\alpha)k_BT\log(f(X))\right),
\label{eq:46}
\end{equation}
where $\Theta$ would be a measure of the surface tension\cite{book:kashchiev-2000,book:kelton-2010}. On the other hand, for 2D nucleation, the work of cluster formation near $X_{min}$ is governed by the line-tension term,
\begin{equation}
\beta\widetilde{\Phi}_{2D}(X)= \beta\left(\Theta X^{1/2}+(1-\alpha)k_BT\log(f(X))\right),
\label{eq:47}
\end{equation}
with $\Theta$ being here the scaled line tension. Hence, we get
\begin{align}
 \int_{X_{min}}^{X_{max}}e^{-\beta\Phi(s)}ds\sim&\int_{0}^{\infty}e^{-\beta\widetilde{\Phi}(s)}ds\label{eq:48},
\end{align}
whose exact results are collected in table \ref{tab:1}. Note that these results involve the Euler gamma function, $\Gamma(n)$.
Besides, 
\begin{align}
 \int_{X_{min}}^{X_{max}}e^{\beta\varphi(s)}ds \sim&\int_0^\infty e^{\beta\left[\varphi(X_*)-\frac{1}{2}|\varphi(X_*)|(s-X_*)^2\right]}ds\notag\\
 \sim&\frac{1}{2}\chi\,e^{\beta\varphi(X_*)},
\label{eq:49}
\end{align}
with 
\begin{align}
 \frac{1}{\lambda}=&\sqrt{\frac{\beta|\varphi''(X_*)|}{2\pi}}=\sqrt{\frac{|\alpha\frac{\delta k_BT}{\nu (X_*)^2}+W''(X_*)|}{2\pi k_BT}}\label{eq:50},\\
 \chi=&\lambda\,\left(1+\mathrm{erf}\left[\frac{\sqrt{\pi}}{\lambda}X_*\right]\right).
\label{eq:51}
\end{align}
The pre-factor $\lambda^{-1}$ turns into the well-known Zeldovich factor,\cite{book:kashchiev-2000} $z_d$, when one selects $\alpha=0$, i.e. It\^{o} calculus. In fact $\chi^{-1}$ could be considered as a generalization of the Zeldovich factor since it plays the same role as the latter in the classical expression of MFPT\cite{book:barrat-hansen-2003-a}:
\begin{equation}
\tau_{_{\text{CNT}}}\sim z_d^{-1}f(X_*)^{-1}e^{\beta W(X_*)}.
\label{eq:52}
\end{equation}
The integral of equation (\ref{eq:48}) depends on the monomer attachment mechanism. In order to perform such an integral we have used the expressions given in the literature for the most usual experimental situations (see table \ref{tab:1}) for both Homogeneous (HON) and Heterogeneous nucleation (HEN):
\begin{itemize}
 \item $f(X)=\zeta X^{1/2}$ for 2D HEN of clusters with monolayer height, with $\zeta$ given by equations (10.6), (10.7), (10.63) and (10.66) of  \citet{book:kashchiev-2000}
 \item $f(X)=\zeta X^{1/3}$ for 3D HON and HEN of caps in liquid or solid solutions, with $\zeta$ given by equations (10.18), (10.20) and (10.24) of \citet{book:kashchiev-2000}
 \item $f(X)=\zeta X^{2/3}$ for 3D HON and HEN of caps when the monomer attachment frequency is controlled by direct impingement or by interface transfer, with $\zeta$ given by equations (10.3), (10.4), (10.5), (10.9), (10.60), (10.61), (10.64) and (10.65) of \citet{book:kashchiev-2000}
\end{itemize}
Hereafter the integral of equation (\ref{eq:48}) will be called $I(\alpha)$. From equation (\ref{eq:45}) we can finally give the approximation of the  MFPT,
\begin{equation}
 \tau(\alpha)\sim \frac{1}{2}\,  \chi\, I(\alpha)\, f(X_*)^{-\alpha}e^{\beta W(X_*)}\label{eq:53},
\end{equation}
which makes it possible the comparison with that predicted by CNT (Eq. \ref{eq:53}),
\begin{align}
\frac{\tau(\alpha)}{\tau_{_{\text{CNT}}}}=&\frac{z_d^{-1}[1+\alpha\,\mathcal{O}\left((X_*)^{-2}\right)]^{-\frac{1}{2}}\,f(X_*)^{-\alpha}I(\alpha)}{z_d^{-1}f(X_*)^{-1}}\nonumber\\
  =&I(\alpha)\,f(X_*)^{1-\alpha}[1+\alpha\,\mathcal{O}\left((X_*)^{-2}\right)]^{-\frac{1}{2}},
\label{eq:54}
\end{align}
which under backward-It\^{o} calculus turns into
\begin{equation}
 \frac{\tau(\alpha=1)}{\tau_{_{\text{CNT}}}}\sim I(1)\label{eq:55}.
\end{equation}
The fact that the approximations are not exactly the same, even though $\alpha=1$ produces the same FPE, it is because we followed a different route to derive $\tau$.

One immediately observes the similitude of equation (\ref{eq:53}) with Kramers' law. In fact, the former is in accordance with the latter with a pre-factor deduced analytically. Yet more, equation (\ref{eq:53}) allows to reach an approximation of the nucleation rate via equation (\ref{eq:40}). Now this result can be extended as a first-order approximation for time-varying conditions,  as shown by \citet{article:getfert-2010} for slowly time-dependent Kramers-Moyal coefficients. Thus, under the assumption of the semiadiabatic limit, and following the notation introduced by \citet{article:getfert-2010}, the instantaneous MFPT can be estimated as follows,
\begin{align}
\tau(\alpha;t)\sim \frac{1}{2}\,  \chi(t)\, I(\alpha,\kappa(t))\, f(X_*,\kappa(t))^{-\alpha}e^{\beta W(X_*,\kappa(t))}.
 \label{eq:56}
\end{align}

While these results can be very useful in order to characterize the time required to start the phase transition, we must bear in mind they are approximations. To get more accurate estimates of this characteristic time an exact numerical integration of the above equations can be performed. Indeed, the best estimation will be determined via stochastic integration of equation (\ref{eq:15}). The main advantage of a numerical approach is that it is valid both for time-independent and time-dependent coefficients.

\begin{table*}[t]
\caption{Integration of equation (\ref{eq:48}) by employing some of the most used expressions for the monomer attachment rate and the corresponding approximations of the MFPT.\label{tab:1}}
\begin{center}
\begin{tabular}{ccccc}
\hline\hline\\
Nucleation type & mass-transport mechanism &$f(X)$ & $\int_{0}^\infty e^{-\beta\widetilde{\Phi}(s)}ds$ & $\tau_{_{\text{MFPT}}}$ \\\\
\hline
&&&&\\
\begin{tabular}{c}
\textbf{2D HEN} \\
{\footnotesize disks with monolayer height}
\end{tabular}&
\begin{tabular}{c}
direct-impingement\footnotemark[1]\\
interface-transfer\footnotemark[2]
\end{tabular}
&
$\zeta_1 X^{1/2} $
& 
$
I_1(\alpha)=\frac{2\Gamma(\alpha+1)\,\zeta_1^{\alpha-1}}{\beta^{\alpha+1}\,\Theta^{\alpha+1}}
$
& $\frac{1}{2}I_1\chi e^{\beta\varphi(X_*)}$\\
&&&&\\
\begin{tabular}{c}
\textbf{3D HON/HEN} \\
{\footnotesize spheres/caps from}\\
{\footnotesize liquid or solid}
\end{tabular}&
\begin{tabular}{c}
volume-diffusion\footnotemark[3]
\end{tabular}
&$\zeta_2 X^{1/3} $& $I_2(\alpha)=\frac{3\,\Gamma\left( \frac{\alpha+2}{2}\right) \,{\zeta_2}^{\alpha-1}}{2\,{\beta}^{\frac{\alpha+2}{2}}\,{\Theta}^{\frac{\alpha+2}{2}}}$& $\frac{1}{2}I_2\chi e^{\beta\varphi(X_*)}$\\
&&&&\\
\begin{tabular}{c}
\textbf{3D HON/HEN} \\
{\footnotesize spheres/caps from vapor,}\\ 
{\footnotesize liquid or solid}
\end{tabular}&
\begin{tabular}{c}
direct-impingement\footnotemark[4]\\
interface-transfer\footnotemark[5]
\end{tabular}
&$\zeta_3 X^{2/3} $& $I_3(\alpha)=\frac{3\,\Gamma\left(\frac{2\,\alpha+1}{2}\right) \,{\zeta_3}^{\alpha-1}}{2\,{\beta}^{\frac{2\,\alpha+1}{2}}\,{\Theta}^{\frac{2\,\alpha+1}{2}}}$ & $\frac{1}{2}I_3\chi e^{\beta\varphi(X_*)}$\\
&&&&\\
\textbf{3D HON/HEN} & arbitrary &$\zeta X^{\delta/\nu}$ & $I(\alpha)=\frac{3\,\Gamma\left( \frac{3}{2}\left( \frac{\alpha\,\delta}{\nu}-\frac{\delta}{\nu}+1\right) \right) \,{\zeta}^{\alpha-1}}{2\,{\beta}^{\frac{3}{2}\left( \frac{\alpha\,\delta}{\nu}-\frac{\delta}{\nu}+1\right)}\,{\Theta}^{\frac{3}{2}\left( \frac{\alpha\,\delta}{\nu}-\frac{\delta}{\nu}+1\right)}} $& $\frac{1}{2}I\chi e^{\beta\varphi(X_*)}$\\
&&&&\\
\hline\hline
\end{tabular}
\end{center}
\footnotetext[1]{$\zeta_1$ prefactor multiplying $N^{1/2}$ in Eqs. (10.6) and (10.7) of Ref. \onlinecite{book:kashchiev-2000}}
\footnotetext[2]{$\zeta_1$  prefactor multiplying $N^{1/2}$ in Eqs. (10.63) and (10.66) of Ref. \onlinecite{book:kashchiev-2000}}
\footnotetext[3]{$\zeta_2$  prefactor multiplying $N^{1/3}$ in Eqs. (10.18), (10.20) and (10.24) of Ref. \onlinecite{book:kashchiev-2000}}
\footnotetext[4]{$\zeta_3$  prefactor multiplying $N^{2/3}$ in Eqs. (10.3)-(10.5) and (10.9) of Ref. \onlinecite{book:kashchiev-2000}}
\footnotetext[5]{$\zeta_3$ prefactor multiplying $N^{2/3}$ in Eqs. (10.60)-(10.65) of Ref. \onlinecite{book:kashchiev-2000}}
\end{table*}

\section{Conclusions\label{conclusions}}

In this work an alternative route to derive classical theory of nucleation has been introduced. Over the course of the last decades, CNT has been exposed to an intense examination which has reported numerous of its flaws and strengths. 
One of the most remarkable strengths of this framework is the majestic simplicity underlying its formulation. However, several different equations for the size distribution came up from such an extensive discussion. Based on the ideas which constitute the classical theory, a unified equation can be reached starting from a unified It\^{o}-Stratonovich stochastic equation.

From a heuristic derivation, the initially unknown coefficients of the proposed stochastic equation were obtained. This stochastic cluster-growth law is interpreted under a general stochastic calculus, despite what is usually done. 
As a first result, our postulated model turns out recovering the cluster-growth law postulated by \citet{article:becker-doring-1935} and \citet{article:tunitskii-1941}, if slowly-varying energy barriers and It\^{o} integration convention are considered. Indeed, such an assumption will be always true within the range of applicability of CNT, since the initial state is near equilibrium. Besides that achievement, we employed the tools of the theory of stochastic processes to go from the stochastic equation to a unified FPE. It is called ``unified'' given the fact that it contains both the classical Zeldovich-Frenkel  equation when we select backward-It\^{o} convention, and the one derived by \citet{article:lutsko-duran-2013}, for the usual Stratonovich calculus. 

Another interesting result generated by this stochastic formalism was found while studying the short-time propagators. They constitute the tools required to know the probability for a cluster to grow or shrink. Surprisingly, when we set the initial cluster to be the critical (whose growth law is supposed to be purely random), one finds that all the interpretations of the noise will give a tendency to grow. It could be argued that this fact would be enough to make us select It\^{o} interpretation. Nevertheless, the only interpretation which has reported the ability to produce a covariant theory has been the Stratonovich calculus\cite{article:lutsko-duran-2013}. Yet more, the study on the general covariance of the theory reported another interesting result, namely Zeldovich-Frenkel equation cannot be fixed to be covariant. This result indeed sheds some light on the question whether or not it is worth trying to fix CNT by considering much more sophisticated models for cluster.

Finally, estimates of the nucleation time and rate were computed. The approximation we reached for these quantities were specifically applied to  the most usual experimental situations. An inevitable similarity to the CNT expressions appeared. Lastly, a comparison with such an expression was performed by computing their ratios.

\section*{Acknowledgements}
M.A.D. acknowledges support from the Spanish Ministry of Science and
Innovation (MICINN), FPI grant BES-2010-038422.
The work of J.F.L is supported in part by the European Space Agency
under Contract No. ESA AO-2004-070 and by FNRS Belgium under Contract No.
C-Net NR/FVH 972. 

\appendix

\section{Short-time propagator\label{appendix:a}}
In order to get the expression of the short-time propagators in a general $\alpha$ convention, we will follow the procedure introduced by  \citet{article:wissel-1979a}, which can also be consulted in \citet{book:risken-1996a}.

The formal solution of a general FPE can be written as (e.g. Ref. \onlinecite{book:risken-1996a})
\begin{equation}
 p(x,t|x',t')=\vec{T}\exp\left[\int_{t'}^t\mathcal{L}_{\text{FP}}[x(s),s]ds\right]\delta(x-x')\label{eq:a-1},
\end{equation}
where $\mathcal{L}_{\text{FP}}$ denotes the Fokker-Planck operator (Eq. \ref{eq:a-3}) and $\vec{T}$ is the time-ordering operator. For small time intervals, $\tau=t-t'$, equation (\ref{eq:a-1}) can be approximated by
\begin{align}
 p(x,&t+\tau|x',t)\notag\\
 =&\ \left[1+\mathcal{L}_{\text{FP}}[x'+\alpha\Delta x,t]\tau+\mathcal{O}(\tau^2)\right]\delta(x-x')\label{eq:a-2},
\end{align}
where we have used the stochastic integral\cite{article:lau-2007} introduced in equation (\ref{eq:2}). Now we can perform the differentiation in the Fokker-Planck operator
\begin{equation}
 \mathcal{L}_{\text{FP}}(u(x),t)=-\frac{\partial}{\partial x} D^{(1)}[u(x),t]+\frac{\partial^2}{\partial x^2} D^{(2)}[u(x),t]
\label{eq:a-3},
\end{equation}
with $u(x)=\alpha x+(1-\alpha)x'$. Accordingly one obtains
\begin{align}
 \mathcal{L}_{\text{FP}}(u(x),t)=&-\alpha\frac{\partial D^{(1)}}{\partial x}(u(x),t)+\alpha^2 \frac{\partial^2 D^{(2)}}{\partial x^2}(u(x),t)\nonumber\\
&-\left[D^{(1)}(u(x),t)-2\alpha\frac{\partial D^{(2)}}{\partial x}(u(x),t)\right]\frac{\partial }{\partial x}\nonumber\\
&+D^{(2)}(u(x),t)\frac{\partial^2}{\partial x^2}\label{eq:a-4}.
\end{align}
By substituting equation (\ref{eq:a-4}) into (\ref{eq:a-2}), and after that using the Taylor expansion of the exponential function we get,
\begin{align}
 p(x,t+\tau|x',t)=&e^{\mathcal{L}_{\text{FP}}(u(x),t)\tau}\delta(x-x').
\end{align}
With the aim of obtaining an exponential function, the Fourier representation of the Dirac-$\delta$ function will be used. Thus,  we finally obtain the sort-time propagator:
\begin{align}
p(x&,t+\tau|x',t)=\notag\\
=&\frac{\exp
\left(
\begin{array}{l}
-\alpha\tau\frac{\partial D^{(1)}(u(x),t)}{\partial x}\\
 \quad \alpha^2\tau\frac{\partial^2D^{(2)}(u(x),t)}{\partial x^2}\\
 \qquad -\frac{\{x-x'-[D^{(1)}(u(x),t)-2\alpha\frac{\partial D^{(2)}}{\partial x}(u(x),t)]\tau\}^2}{4\tau D^{(2)}(u(x),t)}
\end{array}
\right)
}{\sqrt{4\pi\tau D^{(2)}(u(x),t)}}
\label{eq:a-6}.
\end{align}
As can be checked, this short-time propagator recovers those presented by \citet{article:dekker-1976a} and \citet{book:risken-1996a} (e.g. Eqs. (4.55) and (4.55a) of Ref. \onlinecite{book:risken-1996a}) for $\alpha=0$ and $\alpha=1$, respectively. This derivation can be seen as an alternative route to that carried out by \citet{article:lau-2007}.

\bibliographystyle{unsrtnat}
\bibliography{biblio}

\end{document}